\documentclass[preprint,showpacs,preprintnumbers,amsmath,amssymb]{revtex4}
\usepackage{graphicx}
\usepackage{dcolumn}
\usepackage{bm}

\newcommand{\be}{\begin{equation}}
\newcommand{\en}{\end{equation}}
\newcommand{\bea}{\begin{eqnarray}}
\newcommand{\ena}{\end{eqnarray}}

\begin{document}

\title{ Cosmological perturbations in warm inflationary  models with viscous pressure}
\author{Sergio del Campo}
 \email{sdelcamp@ucv.cl}
\affiliation{ Instituto de F\'{\i}sica, Pontificia Universidad
Cat\'{o}lica de Valpara\'{\i}so, Casilla 4059, Valpara\'{\i}so,
Chile.}
\author{Ram\'on Herrera}
\email{ramon.herrera.a@mail.ucv.cl} \affiliation{ Instituto de
F\'{\i}sica, Pontificia Universidad Cat\'{o}lica de
Valpara\'{\i}so, Casilla 4059, Valpara\'{\i}so, Chile.}
\author{Diego Pav\'on}
 \email{diego.pavon@uab.es}
\affiliation{ Departamento  de F\'{\i}sica, Facultad de Ciencias,
Universidad Aut\'onoma de Barcelona, 08193 Bellaterra (Barcelona)
Spain.}

\begin{abstract}
Scalar and tensorial cosmological perturbations generated in warm
inflationary scenarios whose matter-radiation fluid is endowed
with a viscous pressure are considered. Recent observational data
from the WMAP experiment are employed to restrict the parameters
of the model. Although the effect of this pressure on the matter
power spectrum is of the order of a few percent, it may be
detected in  future experiments.
\end{abstract}


\maketitle
\section{Introduction}
As is well known, warm inflation -as opposed to the conventional
``cool" inflation \cite{cool}- has the attractive feature of not
necessitating a reheating phase at the end of the accelerated
expansion thanks to the decay of the inflaton into radiation and
particles during the slow-roll \cite{warm}. Thus, the temperature
of the Universe does not drop dramatically and the Universe can
smoothly proceed into the decelerated, radiation-dominated era
essential for a successful big-bang nucleosynthesis
\cite{peebles}. This scenario has further advantages, namely: (i)
the slow-roll condition $\dot{\phi}^{2} \ll V(\phi)$ can be
satisfied for steeper potentials, (ii) the density perturbations
originated by thermal fluctuations may be larger than those of
quantum origin \cite{origin}, and (iii) it may provide a very
interesting mechanism for baryogenesis \cite{baryo}.

Warm inflation was criticized on the basis that the inflaton
cannot decay during the slow-roll \cite{criticize}. However, in
the recent years, it has been demonstrated that the inflaton can
indeed decay during the slow-roll phase -see \cite{sound} and
references therein- whereby it now rests on solid theoretical
grounds.

Usually, for the sake of simplicity, when studying  the dynamics
of warm inflation the particles created in the decay of the
inflaton are treated as radiation thereby ignoring altogether the
existence of particles with mass in the fluid thus generated.
However, the very existence of these particles necessarily alters
the dynamics as they modify the fluid pressure in two important
ways: (i) its hydrodynamic, equilibrium, pressure is no longer $p
= \rho/3$, with $\rho$ the energy density of the matter-radiation
fluid, but the slightly more general expression $p = (\gamma-1)%
\rho$ where the adiabatic index, $\gamma$, is bounded by $1 \leq
\gamma \leq 2$. (ii) It naturally arises a non-equilibrium,
viscous, pressure $\Pi$, via two different mechanisms: (a) the
inter-particle interactions \cite{lev}, and (b) the decay of
particles within the fluid \cite{yab}.

A radiative fluid (e.g., a mixture of photons and electrons) is a
well-known example of mechanism (a). Actually, it plays a
significant role in the description of the matter-radiation
decoupling in the standard cosmological model
\cite{peebles,weinberg}. Likewise, a hefty viscous pressure arises
in mixtures of different particles species, or of identical
species but with different energies -a case in point is the
Maxwell-Boltzmann gas \cite{MB}.

Concerning mechanism (ii), it is well known that the decay of
particles within a fluid can be formally described by a bulk
viscous pressure, $\Pi$. This is so because the decay is an
entropy-producing scalar phenomenon linked to the spontaneous
widening of the phase space and the bulk viscous pressure is also
an scalar entropy-producing agent. In the case of warm inflation
it has been proposed that the inflaton can excite a heavy field
and trigger the decay of the latter into light fields
\cite{trigger}.

Recently, a detailed analysis of the dynamics of warm
inflation with viscous pressure  showed that when $\Pi%
\neq 0$ the inflationary region takes a larger portion of the
phase space associated to the autonomous system of differential
equations than otherwise \cite{mnp}. It then follows that the
viscous pressure facilitates inflation and lends support to the
warm inflationary scenario.

Our main target is to study the scalar and tensorial cosmological
perturbations associated to this scenario and contrast them with
the data gathered by the the three-year Wilkinson Microwave
Anisotropy Probe (WMAP) \cite{wmap}. We shall follow the method
used in a recent paper by two of us in which the cosmological
perturbations generated by warm inflation driven by a tachyon
field were considered \cite{jcap}. The main differences between
the latter work and the present one are: (i) here the inflaton is
a scalar field, not a tachyon field; (ii) in Ref. \cite{jcap} no
bulk dissipative pressure was considered. For the viscous pressure
we shall assume the usual fluid dynamics expression $\Pi = -3\zeta
H$ \cite{lev}, where $\zeta$ denotes the phenomenological
coefficient of bulk viscosity and $H$ the Hubble function. This
coefficient is a positive-definite quantity (a restriction imposed
by the second law of thermodynamics) and in general it is expected
to depend on the energy density  of the fluid. We shall resort to
the WMAP data to restrict the aforesaid coefficient.

The outline of the paper is as follows. Next Section presents our
model. Sections III and IV deal with the scalar and tensor
perturbations, respectively. Section V specifies to a chaotic
potential and studies the limit of high dissipation. Finally,
Section VI summarizes our findings. We choose units so that $ c=
\hbar =8 \pi G = 1$.

\section{Warm Inflationary model with viscous pressure \label{secti}}
We consider a spatially flat Friedmann-Robertson-Walker (FRW)
universe filled with a self-interacting scalar field $\phi$ (the
inflaton), of energy density, $\rho_\phi=%
\frac{1}{2}\dot{\phi}^{2}+V(\phi)$, and an imperfect fluid, of
energy density $\rho$ and total pressure $p+\Pi$, consisting into
a mixture of matter and radiation of adiabatic index $\gamma$.

The corresponding Friedmann equation reads
 \be
 3H^2\,= \frac{\dot{\phi}^2}{2}+V(\phi)+\rho \, .
 \label{key_02}
 \en
 \\
 Because of the inflaton decays with rate $\Gamma$ into the imperfect
 fluid the conservation equations for the inflaton and the fluid
 generalize to
 \be
 \dot{\rho_\phi}+3H(\rho_\phi+p_\phi)=-\Gamma\dot{\phi}^2\Longrightarrow
\ddot{\phi}+(3H+\Gamma)\dot{\phi}=-V_{\,,\;\phi}\, ,
\label{key_01}
\en
\\
and
 \begin{eqnarray}
 \dot{\rho}+3H(\rho+p+\Pi)=\dot{\rho}+3H(\gamma\rho+ \Pi)=\Gamma\dot{\phi}^2
,\label{3}
\end{eqnarray}
\\
respectively. In general, $\Gamma$ depends on $\phi$ and as a
consequence of the second law of thermodynamics one has  that
$\Gamma= \Gamma(\phi)>0$. As is customary, an upper dot indicates
temporal derivative, and $V_{,\,\phi}=\partial
V(\phi)/\partial\phi$.

During the inflationary phase $V(\phi)$ dominates over any other
form of energy, therefore Friedmann's equation reduces to
\\
\be 3H^{2}= V(\phi) \, .
 \label{inf2} \en
\\
Likewise, imposing the slow-roll conditions, $\dot{\phi}^2 \ll
V(\phi)$, and $\ddot{\phi}\ll (3H+\Gamma)\dot{\phi}$ \cite{cool},
Eq. (\ref{key_01}) becomes
\\
\begin{equation}
3H\left[\,1+r \;\right ] \dot{\phi}=-V_{,\,\phi} \, ,
\label{inf3}
\end{equation}
\\
where the quantity $r \equiv \Gamma/(3H)$ quantifies the
dissipation of the model. Throughout this paper we shall restrict
our analysis to the high dissipation regime, i.e., $r \gg 1$. The
reason for this limitation is the following. Outside this regime
radiation and particles produced both by the decay of the inflaton
and the decay of the heavy fields will be much dispersed  by the
inflationary expansion, whence they will have little chance to
interact and give rise to a non-negligible bulk viscosity.
Likewise, because a much lower number of heavy fields will be
excited the number of decays of heavy fields into lighter ones
will diminish accordingly. (The weak dissipation regime ($r \leq
1$) has been considered by Berera and Fang (second reference of
\cite{warm}) and Moss \cite{Ian1}). Further, if $r$ is not big,
the fluid will be largely diluted and the mean free path of the
particles will become comparable or even larger than the Hubble
horizon. Hence, the regime will no longer be hydrodynamic but
Knudsen's and the hydrodynamic expression $\Pi = - 3 \zeta \, H$
we  are using for the viscous pressure will become invalid. In
such a situation, a consistent analysis should make use of
Boltzmann's equation but this would complicate matters
tremendously. This lies beyond the scope of this paper.

Assuming that the decay of the inflaton is quasi-stable during
inflation it follows that $\dot{\rho}\ll 3 H(\gamma\rho+\Pi)$ and
$ \dot{\rho}\ll\Gamma\dot{\phi}^2$ whence, from Eq. (\ref{3}), we
get
\\
 \begin{equation}
\rho \simeq \frac{r \dot{\phi}^2-\Pi}{\gamma}\label{rh}
\end{equation}
\\
for the energy density of the fluid. It becomes apparent that the
viscous pressure (being necessarily negative) augments the energy
density of the matter-radiation fluid.

With the help of the dimensionless slow-roll parameter
\\
\begin{equation}
\varepsilon\equiv-\frac{\dot{H}}{H^2}=\frac{1}{2(1+r)}
\left[\frac{V_{,\,\phi}}{V}\right]^2\;,\label{ep}
\end{equation}
\\
a useful relation follows between the energy densities, namely,
\\
\begin{equation}
\rho= \frac{1}{\gamma} \left[\frac{2
r}{3(1+r)}\;\varepsilon\;\rho_\phi -\Pi\right]\, .
\label{ro}
\end{equation}

By definition, inflation lasts so long as $\ddot{a} >0$. This
amounts to the condition $\varepsilon<1$, which with the help of
last equation can be expressed as
\\
\begin{equation}
\rho_{\phi}>3\frac{(1+r)}{2 r}\;[\gamma \rho+\Pi]\, .
\label{cond}
\end{equation}
\\
Consequently, warm inflation with viscous pressure comes to a
close when
\\
\begin{equation}
\rho_\phi\simeq 3\frac{(1+r)}{2 r}\;[\gamma \rho+\Pi] .
\end{equation}

The number of e-folds at the end of inflation is given by
\\
\begin{equation}
N(\phi)=-\int_{\phi_i}^{\phi_f}\frac{V}{V_{,\,\phi}}(1+r) \,
d\phi' \, , \label{N}
\end{equation}
\\
where the subscripts $i$ and $f$ stand for  the beginning and end
of inflation, respectively.

For later purpose we introduce here the second slow-roll
parameter, $\eta \equiv-\frac{\ddot{H}}{H \dot{H}}$ \cite{cool},
as a function of the potential and its two first derivatives,
\\
\begin{equation}
\eta\simeq\frac{1}{(1+r)}\left[\frac{V_{,\,\phi\phi}}{V}-\frac{1}{2}
\left(\frac{V_{,\,\phi}}{V}\right)^2\right]. \label{eta}
\end{equation}

\section{Scalar perturbations  \label{pertadiab}}
In terms of the longitudinal gauge, the  perturbed FRW metric can
be written as
\\
\begin{equation}
ds^2=(1+2\Phi)\, dt^{2}-a(t)^{2}(1-2\Psi)\, \delta_{ij}\,
dx^{i}dx^{j},
\end{equation}
\\
where the functions $\Phi=\Phi(t,{\bf x})$ and $\Psi=\Psi(t,{\bf
x})$ denote the gauge-invariant variables of  Bardeen
\cite{Bardeen}. Introducing the Fourier components $e^{i{\bf
kx}}$, with $k$ the wave number, the following set of equations,
in the momentum space, follow  from the perturbed Einstein field
equations -to simplify the writing we omit the subscript $k$-
\\
\begin{equation}
\Phi=\Psi,
\end{equation}
\begin{equation}
\dot\Phi + H\Phi=\frac{1}{2}\;\left[-\frac{(\gamma\rho+\Pi)\; a\;
v}{k}+\dot{\phi}\;\delta\phi \right],
\end{equation}
\\
\begin{equation}
(\delta\phi\ddot{)}+\left[3H+\Gamma\right](\delta\phi\dot{)}+
\left[\frac{k^2}{a^2}
+V_{,\,\phi\phi}+\dot{\phi}\Gamma_{,\,\phi}\right]\;\delta\phi\;\;
=4\dot{\phi}\;\dot{\Phi}-\left[\dot{\phi}\;\Gamma+2V_{,\;\phi}
\right]\;\Phi,
\end{equation}
\\
\begin{equation}
(\delta\rho\dot{)}+3\gamma H\delta\rho+ k a (\gamma \rho+\Pi)
v+3(\gamma
\rho+\Pi)\dot{\Phi}-\dot{\phi}^2\Gamma_{,\,\phi}\delta\phi-\Gamma\dot{\phi}
[2(\delta\phi\dot{)}+\dot{\phi}\Phi]=0,
\end{equation}
\\
and
\\
\begin{equation}
\dot{v}+4H v+\frac{k}{a}\left[\Phi+\frac{\delta
p}{(\rho+p)}+\frac{\Gamma\dot{\phi}}{(\rho+p)}\delta\phi
\right]=0\, ,
\end{equation}
\\
where
\\
\begin{equation}
\delta
p=(\gamma-1)\delta\rho+\delta\Pi\,,\;\;\;\;\;\;
\delta\Pi=\Pi\left[\frac{\zeta_{\,,\,\rho}}{\zeta}\;\delta\rho+\Phi+
\frac{\dot{\Phi}}{H}\right],
\end{equation}
\\
and  the quantity $v$ arises  upon splitting the velocity field as
$\delta u_j =-\frac{i a k_j}{k}\;v\;e^{i {\bf kx}} $ $(j=1,2,3)$
\cite{Bardeen}.

Since the  inflaton and the matter-radiation fluid interact with
each other isocurvature (i.e., entropy) perturbations emerge
alongside the adiabatic ones. This occurs because warm inflation
can be understood as an inflationary model with two basics fields
\cite{Jora1,Jora}. In this context, dissipative effects themselves
can produce a variety of spectral ranging from red to blue
\cite{origin,Jora,62526}, thus producing the running blue to red
spectral suggested by WMAP three-year data \cite{wmap}. We will
come back to this point below.

When looking for non-decreasing adiabatic and isocurvature modes
on large scales, $k\ll a H$ (which depend only weakly on time), it
is permissible to neglect $\dot{\Phi}$ and those terms with
two-times derivatives. Upon doing this and using the slow-roll
conditions, the above equations simplify to
\\
\begin{equation}
\Phi\simeq\frac{1}{2\;H}\left[-\frac{(\gamma\rho+\Pi)\; a\;
v}{k}+\dot{\phi}\;\delta\phi \right]\, ,
\label{PHI}
\end{equation}
\\
\begin{equation}
\left[3H+\Gamma\right](\delta\phi\dot{)}+
\left[V_{,\,\phi\phi}+\dot{\phi}\Gamma_{,\,\phi}\right]\;\delta\phi\;\simeq
-\left[\dot{\phi}\Gamma+2V_{,\;\phi} \right]\;\Phi \, ,
\label{3h}
\end{equation}
\\
\begin{equation}
\delta\rho\simeq\frac{\dot{\phi}^2}{3\gamma
H}[\Gamma_{,\;\phi}\delta\phi+ \Gamma\Phi],\label{r202}
\end{equation}
\\
and
\\
\begin{equation}
 v\simeq
-\frac{k}{4aH}\left[\Phi+\left(\frac{(\gamma-1)\delta\rho+\delta\Pi
}{\gamma\rho+\Pi}\right)+
\frac{\Gamma\dot{\phi}}{\gamma\rho+\Pi}\;\delta\phi\right]\, .
\label{v}
\end{equation}

By combining  of Eqs. (\ref{r202}) and (\ref{v}) with (\ref{PHI}),
the latter becomes
\\
\begin{equation}
\Phi\simeq\frac{\dot{\phi}}{2\;H}\frac{\delta\phi}{G(\phi)}\left[
1+\frac{\Gamma}{4H}+\left([\gamma-1]+\Pi\;\frac{\zeta_{\,,\,\rho}}{\zeta}
\right)\frac{\dot{\phi}\;\Gamma_{\,,\,\phi}}{12\gamma H^2}
 \right]\label{fff},
\end{equation}
where
\begin{equation}
G(\phi)=1-\frac{1}{8\;H^2}\left[
2\gamma\rho+3\Pi+\frac{\gamma\rho+\Pi}{\gamma}\left(\Pi\,\frac{\zeta_{\,,\,\rho}}{\zeta}-1\right)\right].
\end{equation}
\\
In the non-viscous limit, i.e., $\Pi\longrightarrow 0$ when
$G(\phi)\longrightarrow 1$ (since $1\gg \rho/H^2$) and
$\gamma=4/3$, Eq.(\ref{fff}) reduces to the expression obtained
for the conventional warm inflation \cite{Jora1}.

With the help of Eq.(\ref{fff}) and substituting $\phi$ by $t$ as
independent variable, Eq. (\ref{3h}) can be solved. Upon using Eq.
(\ref{inf3}) we get
\\
\begin{equation}
\left(3H+\Gamma\right)\frac{d}{dt}=\left(3H+\Gamma\right)\,\dot{\phi}\frac{d}{d\phi}=
-V_{,\,\phi}\frac{d}{d\phi}\, ,
\end{equation}
\\
and  introducing the ancillary function
\\
\begin{equation}
\varphi=\frac{\delta\phi}{V_{,\,\phi}}\exp\left[\int
\frac{1}{(3H+\Gamma)}\;\Gamma_{,\,\phi}\;d\phi\right],\label{solvar}
\end{equation}
\\
we  are led to the following differential equation for $\varphi$,
\\
\begin{equation}
\frac{\varphi_{,\,\phi}}{\varphi}=-\frac{3}{8\;G(\phi)}\frac{(\Gamma+6H)}{(\Gamma+3H)^2}
\left[\Gamma+4H-\left((\gamma-1)+\Pi\,\frac{\zeta_{\,,\,\rho}}{\zeta}\right)\frac{\Gamma_{,\,\phi}
V_{,\,\phi}}{3\gamma H(3H+\Gamma)}\right]
\,\frac{V_{,\,\phi}}{V}.
\label{var}
\end{equation}

Upon integrating it and resorting to Eq.(\ref{solvar}) we get
\\
\begin{equation}
\delta\phi=C\,V_{,\,\phi} \exp[\Im(\phi)],\label{var2}
\end{equation}
where $\Im(\phi)$ is given by
$$
\hspace{-3.0cm} \Im(\phi)=-\int\left[
\frac{1}{(3H+\Gamma)}\Gamma_{,\,\phi}+\frac{3}{8\;G(\phi)}\frac{(\Gamma+6H)}{(\Gamma+3H)^2}
\left( \Gamma + 4H - \frac{^{}}{_{}} \right. \right.
$$
\begin{equation}
\hspace{5.0cm}\left. \left.
\left[(\gamma-1)+\Pi\,\frac{\zeta_{\,,\,\rho}}{\zeta}\right]\frac{\Gamma_{,\,\phi}
V_{,\,\phi}}{3\gamma
H(3H+\Gamma)}\right)\,\frac{V_{,\,\phi}}{V}\right]d\,\phi,
\end{equation}
and $C$ is an integration constant.

Thus, the density perturbations reads \cite{cool}
\\
\begin{equation}
\delta_{H}=\frac{16\pi}{5}\frac{\exp[-\Im(\phi)]}{V_{,\,\phi}}\,\delta\phi\,.
\label{33}
\end{equation}
\\
It is seen that in the absence of inflaton decay ($\Gamma = 0$,
and therefore, $\Pi=0$),  Eq. (\ref{33}) reduces to the typical
expression for scalar perturbations, $\delta_H\sim%
V\delta\phi/(H\dot{\phi})\sim H\delta\phi/\dot{\phi}$.

For high dissipation ($r \gg 1$), Eq.(\ref{33}) simplifies to
\\
\begin{equation}
\delta_H^2=\left(\frac{16\pi}{15}\right)^2
\frac{\,\exp[-2\;\widetilde{\Im}(\phi)]}{H^2\,r^2\;\dot{\phi}^2}\,\delta\phi^2\,
, \label{331}
\end{equation}
\\
where now $\widetilde{\Im}(\phi):=\Im(\phi)\mid_{r\gg1}$ becomes
\\
\begin{equation}
\widetilde{\Im}(\phi)=-\int\left\{ \frac{1}{3H
r}\Gamma_{,\,\phi}+\frac{3}{8\;G(\phi)}
 \left[1- \left((\gamma-1)+\Pi\,\frac{\zeta_{\,,\,\rho}}{\zeta}\right)
  \frac{\Gamma_{,\,\phi}V_{,\,\phi}}{9\gamma
r\,H^2}\right]\,(\ln(V))_{,\,\phi}\right\}\, d\,\phi \, .
\end{equation}

During the slow-roll phase, the total density fluctuation,
$\delta\rho_T=\delta\rho_\phi+\delta\rho$, and the metric
perturbation, $\Phi$, are related -at first approximation- by
\\
\begin{equation}
\delta\rho_T\simeq
V_{,\,\phi}\,\delta\phi\simeq\,-2[1+r]\,V\,G(\phi)\left[
1+\frac{\Gamma}{4H}+\left([\gamma-1]+\Pi\;\frac{\zeta_{\,,\,\rho}}{\zeta}
\right)\frac{\dot{\phi}\;\Gamma_{\,,\,\phi}}{12\gamma H^2}
\right]^{-1}\,\Phi,\label{r}
\end{equation}
\\
where use of Eq. (\ref{fff}) has been made. Again, for $\Gamma=
\Pi =0$, we recover the usual relation
$\delta\rho_T/\rho_T\simeq\, -2\Phi$, typical of cool inflation,
in which $\rho_{T}=\rho_{\phi}+\rho\simeq V(\phi)$.

In warm inflation, the fluctuations of the scalar field arise
mainly from thermal interaction with the matter-radiation field.
Therefore, following Taylor and Berera  \cite{taylor}, for $r \gg%
1$ we can write
\\
\begin{equation}
(\delta\phi)^2\simeq\,\frac{k_F\,T_r\,}{2\,\pi^2},\label{del}
\end{equation}
\\
where $T_r$ stands for the temperature of the thermal bath and the
wave number $k_F$ is given by $k_F=\sqrt{\Gamma H}=H\,\sqrt{3
r}\geq H$, and corresponds to the freeze-out scale at which
dissipation damps out the thermally excited fluctuations. The
freeze-out wave number $k_F$ is defined at the point where the
inequality $V_{,\,\phi\,\phi}< \Gamma H$, holds \cite{taylor}.

From Eqs. (\ref{33}), (\ref{331}) and (\ref{del}) it follows that
\\
\begin{equation}
\delta^2_H\approx\;\frac{2}{25 \,
\pi^2}\exp[-2\widetilde{\Im}(\phi)]\,
\left[\frac{T_r}{\widetilde{\varepsilon}\,r^{1/2}\,V^{3/2}}\right]
,\label{dd}
\end{equation}
\\
where
$\widetilde{\varepsilon}\approx\frac{1}{2\,r}\left[\frac{V_{,\,\phi}}{V}\right]^{2}$
denotes the dimensionless slow-roll parameter in the high
dissipation phase.

The scalar spectral index $n_{s}$ is given by
\\
\begin{equation}
n_{s} -1 =\frac{d \ln\,\delta^2_H}{d \ln k}
\label{ns}\, ,
\end{equation}
\\
where the interval in wavenumber and the number of e-folds are
related by $d \ln k(\phi)=-d N(\phi)$. Upon using Eqs.(\ref{dd})
and (\ref{ns}), it can be written as
\\
\begin{equation}
n_s  \approx\,
1\,-\,\left[\widetilde{\varepsilon}+2\;\widetilde{\eta}
+\left(\frac{2\widetilde{\varepsilon}}{r}\right)^{1/2}\left[2\widetilde{\Im}_{,\,\phi}-
\frac{r_{,\,\phi}}{2r}\right] \right],\label{ns1}
\end{equation}
where
\\
\begin{equation}
\widetilde{\eta}\approx\frac{1}{r}\left[\frac{V_{,\,\phi\phi}}{V}-\frac{1}{2}
\left(\frac{V_{,\,\phi}}{V}\right)^2\right]
\end{equation}
\\
stands for the second slow-roll parameter, $\eta$, when $r\gg1$.

One interesting feature of the three-year data gathered by the
WMAP experiment is a significant running in the scalar spectral
index $dn_s/d\ln k=\alpha_{s}$ \cite{wmap}. Dissipative effects
can lead to a rich variety of spectral from red to blue
\cite{62526,Jora}. From Eq.(\ref{ns1}) it is seen that in our
model  the running of the scalar spectral index is given by
\\
\begin{equation}
\alpha_s\simeq-\sqrt{\frac{2\;\widetilde{\varepsilon}}{r}}\;\,[\widetilde{\varepsilon}_{\,,\,\phi}
+2\widetilde{\eta}_{\,,\,\phi}]-\frac{\widetilde{\varepsilon}}{r}\,
\left[\left(\frac{\widetilde{\varepsilon}_{\,,\,\phi}}{\widetilde{\varepsilon}}-\frac{r_{\,,\,\phi}}{r}\right)
\left[2\widetilde{\Im}_{,\,\phi}-
\frac{r_{,\,\phi}}{2r}\right]+\left[4\widetilde{\Im}_{,\,\phi\,\phi}-
(\ln(r))_{,\,\phi\phi}\right]\right].\label{dnsdk}
\end{equation}

In models with only scalar fluctuations, the marginalized value of
the derivative of the spectral index can be approximated by
$dn_s/d\ln k=\alpha_s \sim -0.05$ for WMAP-3 only \cite{wmap}.

\section{ Tensor Perturbations\label{GW}}
As argued by Bhattacharya {\it et al.} \cite{bmn}, the generation
of tensor perturbations during inflation gives rise to stimulated
emission in the thermal background of gravitational waves. As a
consequence, an extra temperature dependent factor, $\coth(k/2T)$,
enters the spectrum, $A^2_g\propto\,k^{n_g}$. Thus it now reads,
\\
\begin{equation}
A^{2}_{g}=2\;\left(\frac{H}{2\pi}\right)^2\,\coth\left[\frac{k}{2T}\right]\simeq\frac{V}{6\;\pi^2}
\,\coth\left[\frac{k}{2T}\right]\, , \label{ag}
\end{equation}
\\
the spectral index being
\\
\begin{equation}
n_g=\frac{d}{d\,\ln k}\,\ln\left[
\frac{A^2_g}{\coth[k/2T]}\right]=-2\,\varepsilon \, , \label{ng}
\end{equation}
\\
where we have used Eq.(\ref{ep}).

A quantity of prime interest is the tensor-scalar ratio, defined
as $R(k_{0}) = \left.\left(\frac{A^2_g}{P_{\cal%
R}}\right)\right|_{\,k=k_0}$ where $ P_{\cal R} \equiv 25
\delta_{H}^{2}/4$ and $k_{0}$ is known as the pivot point. Its
expression in the high dissipation limit, $r \gg 1$, follows from
using Eqs. (\ref{dd}) and (\ref{ag}),
\\
\begin{equation}
R(k_0)=\left.\left(\frac{A^2_g}{P_{\cal
R}}\right)\right|_{\,k=k_0}=\left.\,\frac{2}{3}
\left[\left(\frac{\widetilde{\varepsilon}\,r^{1/2}\,V^{5/2}}{T_r}\right)\,\exp[2\,\widetilde{\Im}(\phi)]
\,\coth\left(\frac{k}{2T}\right)\right]\right|_{\,k=k_0}\, .
\label{wg}
\end{equation}

Combining the WMAP three-year data \cite{wmap}  with the SDSS
large scale structure surveys \cite{Teg}, yields the upper bound
$R(k_0$=0.002 Mpc$^{-1}$)$ <0.28$ at $95$\% confidence level,
where $k_0$ = 0.002 Mpc$^{-1}$ corresponds to $\tau_0 \, k_0\simeq
30$, with the distance to the decoupling surface being $\tau_{0} =
14.4$ Mpc. The SDSS is a measure of the galaxy distributions at
red-shifts $a\sim 0.1$ and probes $k$ in the range $0.016\,h$
Mpc$^{-1}$$<k< 0.011\,h\,$Mpc$^{-1}$. The  recent WMAP three-year
data imply for the scalar curvature spectrum the values $P_{\cal
R}(k_0)\equiv\,25\, \delta_H^{2}(k_0)/4\simeq 2.3\times\,10^{-9}$
and the tensor-scalar ratio $R(k_0)=0.095$. We shall make use of
these values to set constraints on the parameters of our model.

\section{The high dissipation limit with chaotic potential \label{exemple}}
In this section we assume for the inflaton the chaotic potential
\\
\begin{equation}
V(\phi)=\frac{1}{2}\, m^{2}\, \phi^{2},\label{pot}
\end{equation}
\\
where $m > 0$ is a free parameter, and (as mentioned above) we
restrict ourselves to study the high dissipation regime ($r \gg
1$).

\subsection{The case $\Gamma=\Gamma_{0}$,  $\zeta = \zeta_{0}$}
When  the inflaton decay rate  and the bulk viscosity coefficient
are given by $\Gamma = \Gamma_{0}=$ constant $>0$ and $\zeta
=\zeta_{0}$= constant, respectively,  the first slow roll
parameter reduces to
\\
\begin{equation}
\widetilde{\varepsilon}=\frac{\sqrt{6}\;m}{\Gamma_0}\frac{1}{\phi}\,
\label{eta1}
\end{equation}
Likewise, the Hubble factor is given by
\\
\begin{equation}
H(\phi)=\frac{m\,\phi}{\sqrt{6}},\label{solH1}
\end{equation}
and the parameter $r$ becomes
\\
\begin{equation}
r=\frac{\sqrt{6}\;\Gamma_0}{3\,m\,\phi}\gg 1 \, .
\end{equation}
\\
Because in this scenario
\\
\begin{equation}
 \dot{\phi}=-\frac{V_{,\,\phi}}{3 r H  },\label{sp2}
\end{equation}
\\
the inflaton depends on time as
\\
\begin{equation}
\phi(t)=\phi_i\;\exp\left[-\frac{m^2\,t}{\Gamma_0}\right]\approx
\phi_i\left[1-\frac{m^2}{\Gamma_0}\;t\right]\label{phisol},
\end{equation}
\\
where $\phi_{i} = \phi(t=t_i=0)$. This entirely agrees with Taylor
and Berera findings \cite{taylor}.

As for the energy density of the matter-radiation fluid  we have
the expression
\\

\begin{equation}
\rho=\frac{m\;\phi}{\gamma}
\left[\frac{\sqrt{6}\;m^2}{3\Gamma_0}+\frac{3\,\zeta_0}{\sqrt{6}}\right]\,,
\end{equation}
\\
or in terms of  $\rho_{\phi}$,
\\
\begin{equation}
\rho=\frac{\sqrt{3}}{\gamma}
\left[\frac{2\;m^2}{3\Gamma_0}+3\,\zeta_0\right]\;\rho_\phi^{1/2}.
\end{equation}

By means of  Eq. (\ref{N}), the number of e-folds  at the end of
warm inflation is found to be
 \\
\begin{equation}
N_{total}=-\int_{\phi_i}^{\phi_f}\,\frac{V}{V_{,\;\phi}}\;r\;d\phi
 =\frac{\sqrt{6}\;\Gamma_{0}}{3\;m}\;[\phi_i-\phi_f],
 \label{N1}
\end{equation}
\\
where, because of $V_{i}>V_{f}$, the initial and final values of
the scalar field satisfy $\phi_{i}>\phi_{f}$.

It is readily seen that $\epsilon_{f}\simeq 1$
 (i.e., $\ddot{a}(t=t_{f})\simeq 0$), and that
the scalar field at the end of inflation reads
\\
\begin{equation}
\phi_f=\frac{\sqrt{6}\;m}{\Gamma_0}.\label{ff}
\end{equation}

Rewriting  the total number of e-folds  in terms of  $\phi_{i}$
and $\phi_{f}$ and  using  Eq.(\ref{ff}), we get
\\
\begin{equation}
\phi_{i}=\frac{1}{2}[
N_{total}+2]\;\phi_f=\frac{\sqrt{6}}{2}\;\frac{m}{\Gamma_0}\;[
N_{total}+2].
\label{N2}
\end{equation}
\\
To comfortably solve the problems of the standard big-bang
cosmology we must have $N_{total}\approx 60$ e-folds. This implies
$\phi_{i}\approx 76 m/\Gamma_{0}$.

At the beginning of inflation
\\
\begin{equation}
r(\phi=\phi_i)=r_i=\frac{1}{93}\;\left(\frac{\Gamma_0}{m}\right)^2\gg
1,
\end{equation}
\\
which implies that $\Gamma_{0}\gg \sqrt{93}\;m$.

From Eq.(\ref{dd}), the scalar power spectrum results to be
\\
\begin{equation}
P_{\cal R}(k_0)\approx \frac{1}{2\pi^2}[8\gamma\Gamma_0
V(\phi_0)^{1/2}+2\sqrt{3}m^2(1-2\gamma)+3\sqrt{3}\zeta_0\Gamma_0(2-3\gamma)]^{3/2}\,
\left[\frac{\Gamma_0^{1/2}\,T_r}{3^{1/4}m^2\,V(\phi_0)^{3/4}}\right],\label{pp}
\end{equation}
\\
Likewise, Eq.(\ref{wg}) provides us with the tensor-scalar ratio
\\
$$
\hspace{-3.0cm}R(k_0)\approx\frac{2}{3}[8\gamma\Gamma_0
V(\phi_0)^{1/2}+2\sqrt{3}m^2(1-2\gamma)+3\sqrt{3}\zeta_0\Gamma_0(2-3\gamma)]^{-3/2}
$$
\begin{equation}
\hspace{9.0cm} \times\,
\left[\frac{3^{1/4}m^2V(\phi_0)^{7/4}}{\Gamma_0^{1/2}
T_r}\right]\coth\left(\frac{k}{2T}\right),\label{raz}
\end{equation}
\\
where $V(\phi_0)$ and  $\phi_0$ stand for the potential and the
scalar field, respectively, when the perturbation, of scale
$k_0=0.002$Mpc$^{-1}$, was leaving the horizon.

By resorting to the WMAP three-year data, $P_{\cal R}(k_0)\simeq
2.3\times 10^{-9}$ and  $R(k_0)=0.095$, and choosing the
parameters $\gamma=1.5$, $m=10^{-6}\,$m$_{P}$, $T\simeq\,T_r\simeq
0.24 \times 10^{16}$ GeV and $k_{0}=0.002\,$Mpc$^{-1}$, it follows
from Eqs. (\ref{pp}) and (\ref{raz}) that $V(\phi_0)\simeq
1.5\times 10^{-11}\,$m$_{P}^{4}$ and $\zeta_0\simeq 3\times
10^{-6}\,$m$_{P}^{3}$. When the scale $k_0$ was leaving the
horizon the inflaton decay rate $\Gamma_{0}$ is seen to be of the
order of $10^{-3}\,$m$_{P}$. Thus Eq. (\ref{dnsdk}) tells us that
one must augment $\zeta_{0}$ by two orders  of magnitude to have a
running spectral index $\alpha_{s}$ close to the observed value
\cite{wmap}.

\subsection{The case $\Gamma = \Gamma (\phi)$, $\zeta = \zeta (\rho)$}
Here we assume, $\Gamma=
\Gamma(\phi)=\alpha\,V(\phi)=\alpha\,m^2\,\phi^2/2$, and
$\zeta=\zeta(\rho)=\zeta_{1}\,\rho$, where $\alpha$ and
$\zeta_{1}$ are positive-definite constants. Obviously,
$\Pi=-3\zeta_{1}\rho\;H$.

Using the chaotic potential, Eq.(\ref{pot}), we find that the slow
roll parameter, the Hubble factor and $r$ can be written as
\\
\begin{equation}
\widetilde{\varepsilon}=\frac{2\sqrt{6}}{m\;\alpha\,\phi^3}\,
,\quad H(\phi)=\frac{m\,\phi}{\sqrt{6}}\, , \quad \mbox{and} \quad
r=\frac{m\,\alpha\,\phi}{\sqrt{6}}\gg 1 \, , \label{solH}
\end{equation}
\\
respectively. By combining the middle equation of (\ref{solH})
with Eq. (\ref{sp2}) we find that
\\
\begin{equation}
\phi^{2}(t)=\phi_i^2\;-\frac{4}{\alpha}\,t \, .\label{phisol2}
\end{equation}

The energy density of the matter-radiation fluid reads
\\
\begin{equation}
\rho=\frac{4\sqrt{3}\;m}{3\phi}
\left[\frac{1}{\sqrt{2}\;\gamma-3\,\zeta_1\,m\,\phi}\right]\,,
\end{equation}
\\
and, in terms of  $\rho_{\phi}$,
\begin{equation}
\rho=\frac{\sqrt{12}\;m^2}{3\rho_\phi^{1/2}}
\left[\frac{1}{\gamma-3\,\zeta_1\,\rho_\phi^{1/2}}\right].
\end{equation}

Using Eq. (\ref{N}), the number of e-folds  at the end of
 warm inflation results to be,
 \\
\begin{equation}
N_{total}=-\int_{\phi_i}^{\phi_f}\,\frac{V}{V_{,\;\phi}}\;r\;d\phi
 =\frac{\alpha\,m\,\sqrt{6}}{36}\;[\phi_i^3-\phi_f^3],
\end{equation}

For $\epsilon_f\simeq 1$ it is seen that the scalar field at the
end of inflation reduces to
\\
\begin{equation}
\phi_f=\left[\frac{2\sqrt{6}}{\alpha\,m} \right]^{1/3} \,
.\label{ff2}
\end{equation}

Rewriting  the total number of e-folds  in terms of  $\phi_f$ and
$\phi_i$, and  using  last equation, we obtain
\begin{equation}
\phi_i=[3\,N_{total}+1]^{1/3}\;\phi_f=[3\,N_{total}+1]^{1/3}\;\left[\frac{2\sqrt{6}}{\alpha\,m}
\right]^{1/3}.\label{N22}
\end{equation}
\\
To get $N_{total}\approx 60$ e-folds we must have $\phi_i\approx
[362\sqrt{6}/(\alpha\, m)]^{1/3}$.

At the beginning of inflation the $r$ parameter becomes
\\
\begin{equation}
r(\phi=\phi_i)=r_i=\left[\frac{181}{3}\right]^{1/3}\;\alpha^{2/3}\;m^{2/3}
\, ,
\end{equation}
\\
resulting in the requirement that $\alpha\gg
\sqrt{\frac{3}{181\,m^2}}$ in the high dissipation regime.

From Eq.(\ref{dd}), the scalar power spectrum is seen to obey
\\
\begin{equation}
P_{\cal R}(k_0)\approx \frac{1}{2\pi^2}\,\Re_0\,
\left[\frac{\alpha^{1/2}\,T_r}{3^{1/4}m^2\,V(\phi_0)^{1/4}}\right]\,
,\label{pp2}
\end{equation}
with
$$
\Re_0= \alpha^2\,V(\phi_0)^2+\beta_1^{16(\gamma+2m^2\zeta_1)/(9\zeta_1-8\gamma)}+%
\left(1+\frac{\beta_2}{\beta_3}\right)^{\beta_4}%
-\left(1-\frac{\beta_2}{\beta_3}\right)^{\beta_4}\, ,
$$
being $\zeta_{1} \neq 8\gamma/9$, and where
$$
\beta_1=-8\gamma
V(\phi_0)+6\gamma^2-12\sqrt{3}\gamma\zeta_1\sqrt{V(\phi_0)}-3\gamma+3\sqrt{3}\zeta_1\sqrt{V(\phi_0)}+9\zeta_{1}^{2}
V(\phi_0),
$$
$$
\beta_2=2\sqrt{V(\phi_0)}(8\gamma-9\zeta_1^2)+3\sqrt{3}\zeta_1(4\gamma-1),
$$
$$
\beta_3=(8\gamma-9\zeta_1^2)\sqrt{
192\gamma^3+216\gamma^2\zeta_1^2-96\gamma^2-108\gamma\zeta_1^2+27\zeta_1^2
}\;,
$$
and
$$
\beta_4=\left(\frac{8\sqrt{3}}{3\beta_3}\right)[64m^2\gamma^2+27m^2\gamma\zeta_1^2
-64m^2\gamma+36m^2\zeta_1+72\zeta_1\gamma^2-18\zeta_{1}\, \gamma].
$$

From Eq.(\ref{wg}) the tensor-scalar ratio can be obtained as
\\
\begin{equation}
R(k_0)\approx\frac{2}{3}\,
\left[\frac{3^{1/4}\;m^2\;V(\phi_0)^{5/4}}{\alpha^{1/2}\;\Re_0\;
T_r}\right]\coth\left(\frac{k}{2T}\right)\, . \label{raz2}
\end{equation}

Resorting again to the WMAP three-year data ($P_{\cal
R}(k_0)\simeq 2.3\times 10^{-9}$, $R(k_0)=0.095$) and choosing the
parameters $\gamma=1.5$, $\alpha\simeq 10^7 \,$m$_{P}^{-3}$,
$V(\phi_0)\simeq 1.5\times 10^{-11}\,$m$_{P}^{4}$,
$T\simeq\,T_r\simeq 0.24 \times 10^{14}$ GeV and
$k_0=0.002$Mpc$^{-1}$, we find from Eqs. (\ref{pp2}) and
(\ref{raz2}) that $m\simeq 10^{3}\,$m$_{P}$,  and $\zeta_1\simeq
10^{-8}\,$m$_{P}^{-1}$. We note that the dissipation coefficient
when the scale $k_0$ was leaving the horizon is of the order of
$\Gamma(\phi_0)=\alpha\,V(\phi_0)\sim 10^{-4}\,$m$_{P}$. It
follows from Eq.(\ref{dnsdk}) that one must decrease $\zeta_{1}$
by three  orders of magnitude to have a running spectral index
$\alpha_s$ close to the WMAP observed value.

\section{Concluding remarks \label{conclu}}
In this paper we considered a warm inflationary scenario in which
a viscous pressure is present in the matter-radiation fluid. This
pressure arises on very general grounds, either as a
hydrodynamical effect \cite{lev} or as a consequence of the decay
of massive fields -previously excited by the inflaton- into light
fields, or both. We investigated the corresponding scalar and
tensor perturbations. The contributions of the adiabatic and
entropy modes were obtained explicitly. Specifically, a general
relation for the density perturbations is given in Eq.(\ref{33}).
The tensor perturbations are generated via stimulated emission
into the existing thermal background, Eq. (\ref{ag}), and the
tensor-scalar ratio -as well as the dissipation parameter- is
modified by a temperature dependent factor.

The effect of the viscous pressure reveals itself at the e-folding
level. Indeed, in the first case, i.e., when $\Gamma= \Gamma_{0}=$
constant and $\zeta = \zeta_{0}$, as Eq.(\ref{N1}) shows,  the
total number of e-folds depends on the difference of the inflaton
field evaluated at the beginning and at the end of inflation,
i.e., $\phi_{i}-\phi_{f}$, where the initial value of the inflaton
field is $ \phi_{i}^{2} = \frac{3%
H_{i}}{m^{4}\,\Gamma_{0}}(3H_{i}+\Gamma_0)^{2}(\gamma\,\rho_{i}+\Pi_{i})$.
Because $\Pi_{i }< 0$ one has that $\phi_{i}$ is lower than when
the viscous pressure is not considered, thereby the viscous
pressure shortens the inflationary phase.

Also, in this first case, we constrained the parameters of our
model with the help of the WMAP three-year data \cite{wmap}. Thus,
we have found that $\zeta_{0}$ should be the order of
$10^{-6}\,$m$_{P}^{3}$, and the potential, $ V(\phi_0)$, of the
order of $10^{-11}\,$m$_{P}^{4}$ when the perturbation exits the
horizon at the scale of $k_0=0.002\,$Mpc$^{-1}$. Likewise, we have
found that the contribution of the viscous pressure to the scalar
power spectrum, i.e., $ \chi \equiv \frac{P_{\cal R} -P_{\cal%
R}^{(\Pi= 0)}}{P_{\cal R}}$, is the order of two percent.

Similar results follow in the second case, i.e., when $\Gamma =
\Gamma(\phi)$ and $\zeta = \zeta(\rho)$. In this instance, the
ratio $\chi$ is about four percent. While the deviation in both
cases is small (but not negligible), we think it is still big
enough to be detected by future measurements of the large scale
structure of the Universe.

We have implicitly assumed that $\gamma$ is a constant which is
not really the case since matter and radiation redshift at
different rates, and that the bulk viscosity pressure is given by
the conventional stationary formula $\Pi = -3 \zeta H$ while it is
well known that it does not respect relativistic causality and
should be generalized to the Israel-Stewart expression $\tau
\dot{\Pi}+ \Pi = -3 \zeta H$, where $\tau$ is the relaxation time
of the viscous process \cite{werner}. However, it is to be
expected that these generalizations, when properly incorporated,
will not substantially alter our findings.

We have not studied the effect of the viscous pressure on the
bispectrum of density perturbations -though there are good reasons
to do it- as it lies a bit outside the main target of our paper.
While cool inflation typically predicts a nearly vanishing
bispectrum, and hence a small (just a few per cent) deviation from
Gaussianity in density fluctuations -see e.g. \cite{vanishing}-,
warm inflation clearly predicts a non-vanishing bispectrum. The
latter effect arises from the non-linear coupling between the the
fluctuations of the inflaton and those of the radiation. This can
produce  a moderate non-Gaussianity \cite{moderate} or even a
stronger one -likely to be detected by the Planck satellite
\cite{mission}- if the aforesaid nonlinear coupling is extended to
subhorizon scales \cite{Ian2}. Because $\Pi$ implies an additional
coupling between the radiation and density fluctuations it is to
be expected that non-Gaussianity will be further enhanced.
Perhaps, this could serve to observationally constrain $\Pi$ by
future experiments. We plan to consider this question in a
subsequent paper.

In summary, our model  presents two interesting features: (i)
Related to the fact that the dissipative effects plays a crucial
role in producing the entropy mode, they can themselves produce a
rich variety of spectral ranging from red to blue. The possibility
of a spectrum which does run so is particularly interesting
because it is not commonly seen in inflationary models which
typically predict red spectral. (ii) The viscous pressure may tell
us about how the matter-radiation component behaves during warm
inflation. Specifically, it will be very interesting to know how
the viscosity contributes to the large scale structure of the
Universe. In this respect, we anticipate that the Planck mission
\cite{mission} will significantly enhance our understanding of the
large scale structure by providing us with high quality
measurements of the fundamental power spectrum over an larger
wavelength range than the WMAP  experiment.

\begin{acknowledgments}
SdC and RH wish to thank the ``Universidad Aut\'{o}noma de
Barcelona" for hospitality where part of this work was done. SdC
was supported by Comision Nacional de Ciencias y Tecnolog\'{\i}a
through FONDECYT grants N$^0$ 1030469, and ``FONDECYT-Concurso
incentivo a la cooperaci\'on internacional"  N$^0$ 7060005. Also,
from UCV-DGIP N$^0$ 123.764. RH was supported by the ``Programa
Bicentenario de Ciencia y Tecnolog\'{\i}a" through the Grant
``Inserci\'on de Investigadores Postdoctorales en la Academia"
\mbox {N$^0$ PSD/06}. DP acknowledge ``FONDECYT-Concurso incentivo
a la cooperaci\'on internacional"  N$^0$7060005, and is grateful
to the ``Instituto de F\'{\i}sica" de la PUCV for warm
hospitality, also DP research's was partially supported by the
``Ministerio Espa\~{n}ol de Educaci\'{o}n y Ciencia" under Grant
FIS2006-12296-C02-01.
\end{acknowledgments}

\end{document}